\documentclass[%
 aip,
 amsmath,amssymb,
 reprint,%
]{revtex4-1}

\pdfoutput=1

\usepackage{graphicx}
\usepackage{epstopdf}
\usepackage{dcolumn}
\usepackage{bm}
\usepackage{lipsum,mwe}
\usepackage[utf8]{inputenc}
\usepackage[T1]{fontenc}
\usepackage{mathptmx}

\usepackage{xcolor}
\usepackage{color}
\usepackage{upgreek}

\newcommand{\wcm}{\,\mathrm{W/cm}^2}

\newcommand{\micron}{\,\upmu\mathrm{m}}

\begin{document}

\preprint{AIP/123-QED}

\title[]{Interaction of ultraintense radially-polarized laser pulses with plasma mirrors}

\author{N. Zaïm}%
\thanks{These three authors contributed equally}
\affiliation{ LOA, CNRS, \'Ecole Polytechnique, ENSTA Paris, Institut Polytechnique de Paris, 181 Chemin de la Hunière et des Joncherettes, 91120 Palaiseau, France}
\author{D. Gu\'enot}
\thanks{These three authors contributed equally}
\affiliation{ LOA, CNRS, \'Ecole Polytechnique, ENSTA Paris, Institut Polytechnique de Paris, 181 Chemin de la Hunière et des Joncherettes, 91120 Palaiseau, France}
 \affiliation{ Department of physics, Lund University, SE-22100 Lund, Sweden}
\author{L. Chopineau}%
\thanks{These three authors contributed equally}
\affiliation{ Lasers, Interactions and Dynamics Laboratory (LIDyL), Commissariat \`{a} l'\'Energie Atomique, Universit\'e Paris-Saclay, DSM/IRAMIS, CEN Saclay, 91191 Gif sur Yvette, France}
\author{A. Denoeud}%
\affiliation{ Lasers, Interactions and Dynamics Laboratory (LIDyL), Commissariat \`{a} l'\'Energie Atomique, Universit\'e Paris-Saclay, DSM/IRAMIS, CEN Saclay, 91191 Gif sur Yvette, France}
\author{O. Lundh}%
\affiliation{ Department of physics, Lund University, SE-22100 Lund, Sweden}
\author{H. Vincenti}%
\affiliation{ Lasers, Interactions and Dynamics Laboratory (LIDyL), Commissariat \`{a} l'\'Energie Atomique, Universit\'e Paris-Saclay, DSM/IRAMIS, CEN Saclay, 91191 Gif sur Yvette, France}
\author{F. Qu\'er\'e}%
\affiliation{ Lasers, Interactions and Dynamics Laboratory (LIDyL), Commissariat \`{a} l'\'Energie Atomique, Universit\'e Paris-Saclay, DSM/IRAMIS, CEN Saclay, 91191 Gif sur Yvette, France}
\author{J. Faure}%
\email{jerome.faure@ensta-paris.fr}
\affiliation{ LOA, CNRS, \'Ecole Polytechnique, ENSTA Paris, Institut Polytechnique de Paris, 181 Chemin de la Hunière et des Joncherettes, 91120 Palaiseau, France}

\date{\today}

\begin{abstract}
We present experimental results of vacuum laser acceleration (VLA) of electrons using radially polarized laser pulses interacting with a plasma mirror. Tightly focused radially polarized laser pulses have been proposed for electron acceleration because of their strong longitudinal electric field, making them ideal for VLA. However, experimental results have been limited until now because injecting electrons into the laser field has remained a considerable challenge. Here, we demonstrate experimentally that using a plasma mirror as an injector solves this problem and permits to inject electrons at the ideal phase of the laser, resulting in the acceleration of electrons along the laser propagation direction while reducing the electron beam divergence compared to the linear polarization case. We obtain electron bunches with few-MeV energies and a 200 pC charge, thus demonstrating for the first time electron acceleration to relativistic energies using a radially polarized laser. High-harmonic generation from the plasma surface is also measured and provides additional insight into the injection of electrons into the laser field upon its reflection on the plasma mirror. Detailed comparisons between experimental results and full 3D simulations unravel the complex physics of electron injection and acceleration in this new regime: we find that electrons are injected into the radially polarized pulse in the form of two spatially-separated bunches emitted from the p-polarized regions of the focus. Finally, we leverage on the insight brought by this study to propose and validate a more optimal experimental configuration that can lead to extremely peaked electron angular distributions and higher energy beams.
\end{abstract}

\maketitle

\section{\label{sec:level1} Introduction}
Owing to the progress of intense femtosecond lasers~\cite{Yu12}, new methods for accelerating particles have been developed in the last two decades. Most notably, laser wakefield accelerators~\cite{Tajima79,Esarey09} in underdense plasmas take advantage of extremely high accelerating gradients, on the order of 100 GV/m, to generate ultrashort~\cite{lund11} electron bunches with high beam quality and energies ranging from few-MeV~\cite{guen17} to multi-GeV~\cite{gons18}. Vacuum Laser Acceleration (VLA), in which the electrons are directly accelerated by the laser field in vacuum, is another method for accelerating electrons that has also drawn significant attention. Many theoretical~\cite{hart95,esar95b,Quesnel98,yu00,dodi03} and numerical~\cite{stu01,sal02,pang02,malt03} studies of VLA have been carried out with the prospects of understanding this fundamental interaction and profiting from the immense accelerating fields involved, that can exceed 10 TV/m. However, experimental observation of high energy gains from VLA has proven difficult to achieve~\cite{moor99,paye12,clin13,carb16}. This is due to the fact that acceleration is efficient when electrons are injected with a high initial velocity at a precise phase of the laser pulse, close to a zero of the electric field, so that they remain in an accelerating phase of the laser for a long time. Since the accelerating structure has superluminal phase velocity~\cite{esar95b} and usually sub-micron wavelength, these stringent injection conditions are not attained with conventional injection methods. Indeed, two methods have been tried so far to inject electrons into an intense laser field: the ionization of a low density gas target~\cite{moor99,paye12} or the use of a pre-accelerated electron beam~\cite{clin13,carb16}. In the first case, electrons are injected with no initial velocity at a phase of the laser that is not optimal for electron acceleration (close to a maximum of the electric field), which leads to inefficient acceleration. In the second case it is particularly challenging to inject the electrons at a precise phase of the accelerating structure, as it would require electron bunches with attosecond duration and synchronization with the laser. For this reason, experimental attempts at VLA with this method tend to result in a widening of the energy spread rather than a net acceleration~\cite{clin13,carb16}.

It was recently observed that plasma mirrors, i.e. overdense plasmas with a sharp density gradient on their front surface ($L \ll \lambda_0$, where $L$ is the gradient scale length and $\lambda_0$ the laser wavelength), could solve this issue and act as ideal injectors for VLA~\cite{thevenet16}. Indeed, when a p-polarized laser pulse with ultrahigh intensity is focused on such target, the laser makes the plasma mirror surface oscillate at the laser frequency and at relativistic velocities. These oscillations lead to the periodic emission of high-harmonics, via the Relativistic Oscillating Mirror (ROM) mechanism~\cite{Thaury10}, and ejection of electrons at a precise phase of the laser pulse~\cite{Thevenet16_2}. These electrons are ideally injected in the reflected laser field: they begin their interaction close to a zero of the electric field with a relativistic velocity directed towards the specular direction. This allows them to subsequently gain large amounts of energy in vacuum from the reflected pulse. Using 20\,TW laser pulses on-target, it was demonstrated that 3\,nC, 5-10\,MeV electron bunches with 150\,mrad divergence could be produced~\cite{thevenet16,chop19}.

When VLA is carried out with linearly polarized pulses, the accelerating fields are transverse. This means that the electrons are pushed off the optical axis as they are accelerated, which tends to widen their angular distribution. For this reason, the possibility of accelerating electrons with longitudinal electric fields has frequently been studied~\cite{esar95b,sala06,karm07,vari13,marc13,mart14,wong17}. Radially polarized beams are ideal candidates for achieving this~\cite{sala06,karm07,vari13,marc13,mart14,wong17}. A laser is radially polarized if at every position the polarization vector points towards its center. The relevance of this polarization state comes from the presence of a strong longitudinal electric field at focus, which is maximum on the optical axis, where the transverse fields vanish. The main components of a paraxial radially polarized pulse at focus can be written in cylindrical coordinates as:
\begin{gather}
E_r = E_{0,z}\frac{k_0 r}{2} \exp\left(-\frac{r^2}{w_0^2}\right)\cos{\omega_0 t} \times f(t)
\\
B_{\theta} = E_r/c
\\
E_z = E_{0,z}\exp\left(-\frac{r^2}{w_0^2}\right)\left(1-\frac{r^2}{w_0^2}\right)\sin{\omega_0 t} \times f(t).
\end{gather}
where $E_{0,z}$ is the peak amplitude of the longitudinal electric field, $c$ is the speed of light in vacuum, $k_0 = 2\pi / \lambda_0$ is the laser central wavenumber, $\omega_0$ its central frequency, $w_0$ its beam waist and $f(t)$ a normalized temporal envelope. For a radially polarized beam, the magnetic field only has an azimuthal component $B_\theta$. The transverse fields ($E_r$ and $B_\theta$) fade on the optical axis, resulting in doughnut-shaped spatial intensity profiles.

It is common to define $a_{0,z} = E_{0,z}/E_0$ as the normalized amplitude of the longitudinal field. Here $E_0 = m_e c \omega_0 / e$ where $m_e$ and $e$ are the electron mass and charge respectively. The normalized peak amplitude of the radial field, noted $a_{0,r}$, is given by $a_{0,r} = \frac{\exp ( -0.5 )}{\sqrt{2}} k_0 w_0 a_{0,z}$. It can be noted that $a_{0,z}/a_{0,r} \propto 1/w_0$. Therefore, the longitudinal field is predominant only in the case of very tight focusing. If the $E_z$ field is strong enough, it can be used to directly accelerate electrons in the longitudinal direction. This requires the electrons to be located very close to the optical axis, where the longitudinal field is maximum and the deflecting $E_r$ and $B_\theta$ fields are negligible. In such a case, the electrons can remain for a long time in the center of the laser beam, which can in principle result in an electron beam with higher energy and smaller divergence than with linear polarization~\cite{karm07}. In the past, there have been two experimental attempts at VLA using radial polarization, but modest energy gains, of at most tens of keV, have been achieved~\cite{paye12,carb16}. This is due to the fact that relatively small laser energies were used, and, as explained earlier, that electron injection into the laser pulse was not optimal. However, a recent study~\cite{Zaim17} has demonstrated that, as in the case of linear polarization, plasma mirrors could be used to inject electrons with ideal initial conditions into radially polarized beams, resulting in relativistic electron beams accelerated by the $E_z$ field.

This paper presents experimental results of electron acceleration to relativistic energies using radially polarized pulses combined to a plasma mirror injector. We observe that using radial polarization leads to electron acceleration in the longitudinal direction, thereby reducing the angular spread. Electron beams with a charge exceeding 100~pC and an energy spectrum peaking at 2 MeV have been obtained. In order to gain further insight from the interaction and the electron injection process, we also study the case of azimuthal polarization. Azimuthally polarized beams are similar to radially polarized beams but have the electric and magnetic fields "exchanged", meaning that the main components of such pulses are $E_\theta$, $B_r$ and $B_z$. They do not exhibit an $E_z$ field and therefore cannot directly accelerate electrons in the longitudinal direction, which we verify experimentally. We additionally present full-3D Particle-In-Cell (PIC) simulations that shed light on the intricate physics of this interaction and eventually allows us to propose an experimental configuration which is more optimal for VLA with radial polarization. While the propagation of ultraintense radially or azimuthally polarized pulses in underdense plasmas has been previously studied~\cite{naka16}, this is the first time that the interaction of such pulses with overdense plasmas is investigated experimentally. 

\begin{figure*}[ht!]
  \includegraphics[width=2.\columnwidth,scale=1]{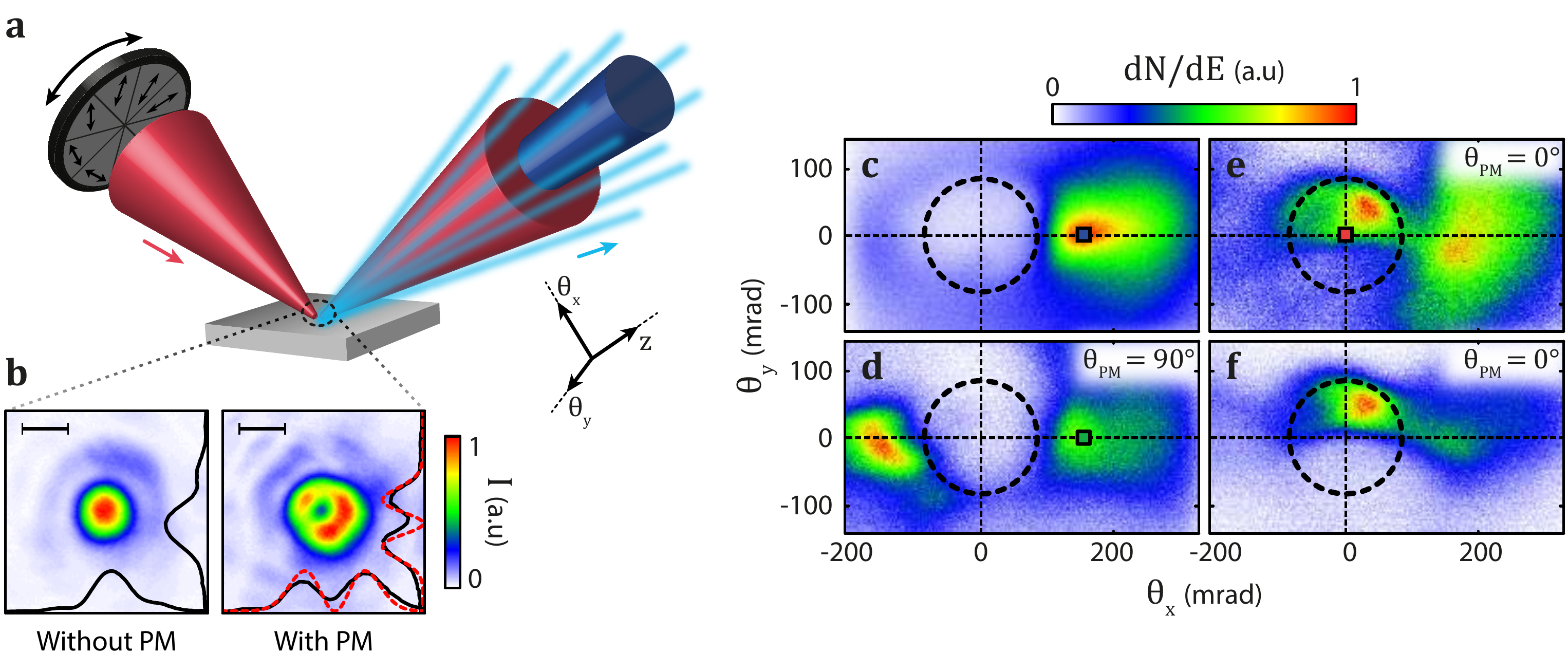}
  \caption{\label{fig:Setup} (a) Sketch of the experiment. An ultraintense laser pulse (red) is reflected on the target (gray). Relativistic electrons (light blue) and high-harmonics (dark blue) are generated from the interaction and are measured simultaneously. A phase mask (PM) made of eight octants, depicted in the top left corner, can be inserted to convert the laser polarization from linear to radial. By rotating the phase mask, the polarization can then be continuously varied from radial to azimuthal. (b) The left panel shows a focal spot for linear polarization, i.e. without the phase mask, while the right panel shows a focal spot for radial polarization. (c)-(f) Typical experimental angular electron distributions. The specular direction corresponds to $\theta_x=\theta_y=0 $. (c) Linear polarization: the electron beam is located between the specular and normal directions. (d) Azimuthal polarization ($\theta_{PM}=90^{\circ} $): electrons are located on both sides of the specular direction. (e)-(f) Radial polarization ($\theta_{PM}=0^{\circ} $): Panel (e) shows a typical angular distribution while panel (f) is a result from one of the best shots. Both distributions display an electron beam in the specular direction. The dashed black circles show the angular extent of the reflected laser beam.}
\end{figure*}

The paper is organized as follows. In Sec.~\ref{sec:level2}, we describe our experimental setup. The experimental results are presented in Sec.~\ref{sec:level3} and analyzed in Sec.~\ref{sec:level4} with the help of full-3D Particle-In-Cell (PIC) simulations. Finally, the ideal parameters that would be required to optimize the acceleration with radially polarized beams are discussed in Sec.~\ref{sec:level5}.

\section{\label{sec:level2} Experimental setup}
The experiment is represented in Fig.~\ref{fig:Setup}(a). The UHI100 laser at CEA Saclay is a 100-TW class system which provides 800-nm, 25-fs laser pulses with an ultrahigh temporal contrast ($>10^{12}$ up to 10 ps before the main pulse) thanks to a double plasma mirror system~\cite{Levy07} located before the experimental chamber. We use a deformable mirror to correct the laser wavefront. A phase mask consisting of eight half-wave plates with different optical axes, pictured in the top-left corner of Fig.~\ref{fig:Setup}(a), can be inserted in order to convert the laser polarization from linear to radial or azimuthal. Each octant of the phase mask is made of an 80-$\micron$ thick piece of mica, which is thin enough to result in a low B-integral of 0.15 rad. By rotating the entire phase mask, the polarization can be continuously varied from radial to azimuthal. A circular aperture is used to remove the edges of the beam and improve the focal spot. In the case of linear polarization, a $50$\,mm aperture is used, thus reducing the energy on target to $460$\,mJ. When the radial polarization phase mask is introduced in the beam, a $65$\,mm aperture is used such that the energy on target is $675$\,mJ. The beam is focused with a $60^\circ$ incidence angle onto a fused silica target by a parabola with a focal length $f = 200$~mm. Resulting focal spots are shown in Fig.~\ref{fig:Setup}(b). For linear polarization, the beam waist is measured to be $3\times3.4\,\upmu$m, which results in an estimated peak intensity of $I = 5.8\times10^{19} \wcm$ ($a_0 = 5.2$). For radial polarization, the characteristic doughnut shape is clearly observed. The slight asymmetry of the spot is probably due to imperfections in the wavefront and/or imperfect centering of the beam on the phase mask. The beam waist is obtained by fitting $r^2 \exp{\left( -2r^2/w_0^2 \right)} $ to the spatial intensity profile, which gives $3.05 \times 3.2\,\upmu$m. The resulting peak intensity is $I = 4.8\times10^{19} \wcm$ ($a_{0,r} = 4.7$). Using these parameters we estimate the longitudinal normalized field to be: 
\begin{align}
a_{0,z} = 0.742 a_{0,r}\left(\frac{\lambda_0}{w_0}\right) = 0.9.
\end{align}
This value is probably somewhat overestimated because of the asymmetry in the focal spot but it indicates that the longitudinal component of the electric field approaches relativistic intensities, making the laser suitable for electron acceleration in the longitudinal direction.

A small mirror is inserted before the parabola in order to create a weak pre-pulse from the main laser beam~\cite{Kahaly13}. This pre-pulse is used to ionize the target and initiate a plasma expansion at an adjustable delay before the main pulse, therefore creating a transversely homogeneous preplasma with an accurately controlled density gradient.
 
A calibrated~\cite{Glinec06} phosphor screen (KODAK LANEX fine) is used in combination with a camera to observe the electron angular distribution around the specular direction. The screen is covered by a 1.63 mm aluminium plate to provide shielding against the laser light and remove the electrons with an energy lower than $\sim$900 keV. A pair of magnets in combination with a slit can be added in front of the screen to measure angularly resolved electron energy spectra in the incidence plane ($\theta_y = 0$, with $\theta_y$ defined in Fig.~\ref{fig:Setup}(a))~\cite{chop19}.

In parallel to the electron distributions, high order harmonics emitted in the specular direction can be measured by replacing the electron spectrometer with an XUV spectrometer. The spectrometer uses a 1200 lines per mm varied line spacing concave extreme ultraviolet grating (Shimadzu 30-002) coupled to a large $69 \times 88$\,mm rectangular microchannel plate and a phosphor screen. This provides harmonic spectra which are angularly resolved in the transverse $\theta_y$ direction. With linear polarization, we observe harmonics up to the $45^{th}$ order approximately.

\section{\label{sec:level3} Experimental Results}

A typical electron angular distribution obtained with linear polarization is displayed in Fig.~\ref{fig:Setup}(c). The main features of this distribution are the presence of a hole in the specular direction and of a bright peak between the specular and the normal directions, in good agreement with previous experiments~\cite{thevenet16,chop19}. This bright spot is located approximately $150$\,mrad from the specular direction and contains a charge of $\approx 700$\,pC. The electron signal in this regime is optimal when the gradient scale length is on the order of $\lambda_0/10$ (see Ref.~\cite{chop19}).

\begin{figure}[h]
\centering
\includegraphics[width=1.\columnwidth,scale=1]{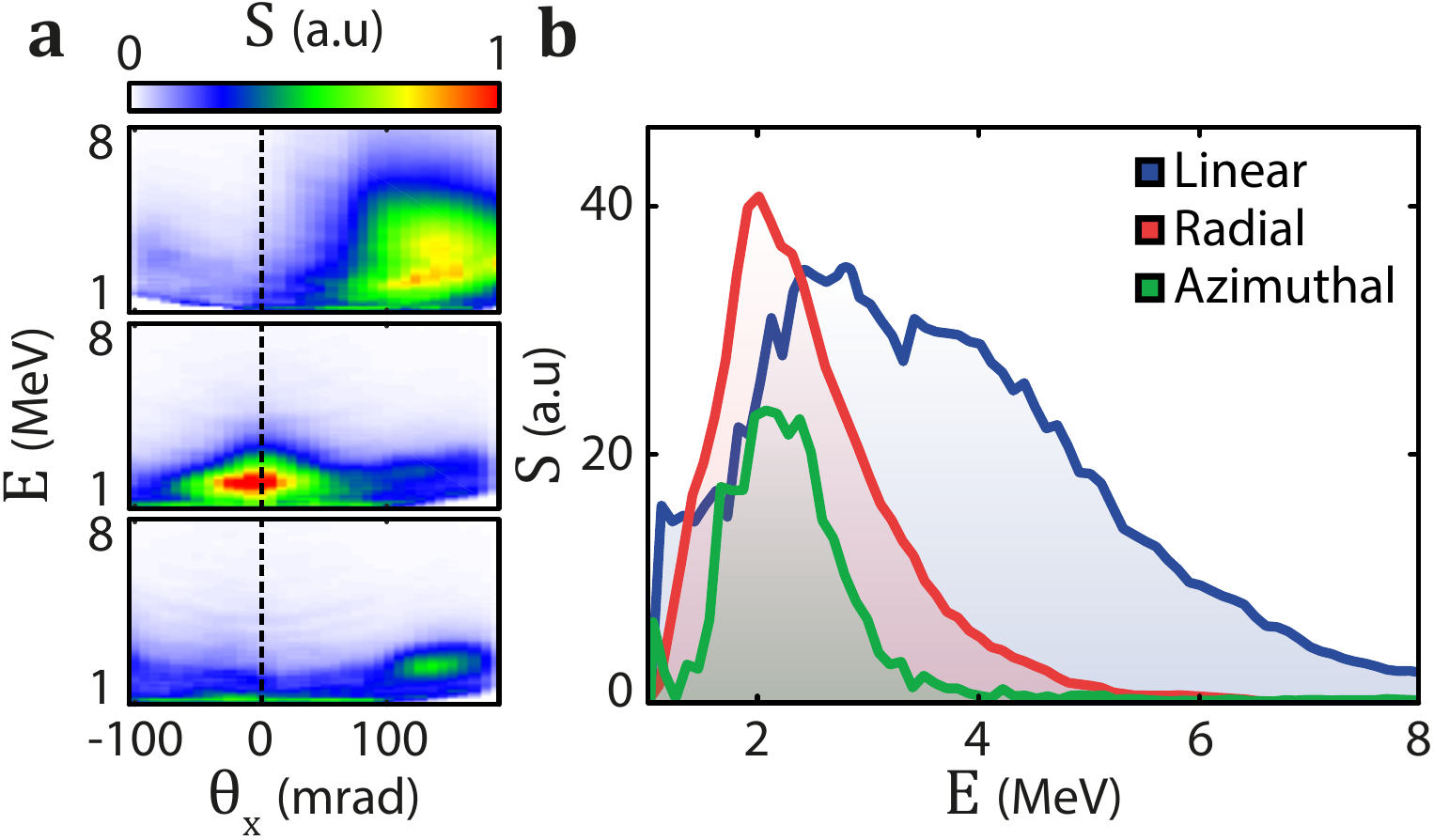}
\caption{\label{fig:spectra} (a) Angularly resolved energy spectra of the electrons emitted near the incidence plane ($\theta_y = 0$) for linear polarization (top panel), radial polarization (middle panel) and azimuthal polarization (bottom panel). The dashed line marks the specular direction. (b) Electron energy spectra obtained at the position of the square markers in Fig.~\ref{fig:Setup}(c)-(f) for linear polarization (blue curve), radial polarization (red curve) and azimuthal polarization (green curve).}
\end{figure}

While the electron signal exhibits good stability in linear polarization, we observe significant shot-to-shot fluctuations in the radial polarization case, which we attribute to shot-to-shot fluctuations in the laser focal spot that appear when the phase mask is inserted in the laser beam. We nonetheless consistently observe an electron beam emitted very close to the specular direction, while another spot remains visible between the normal and specular directions, as can be seen in Fig.~\ref{fig:Setup}(e). This peak in the specular direction has a smaller divergence, typically in the $50$\,mrad range, and can contain up to $200$\,pC ($100$\,pC on average). For the best shots, the spot in the specular direction can even contain more charge than the spot located between specular and normal, and the electron beam divergence is reduced by a factor of two compared to the linear polarization case. An example of such shot is shown in Fig.~\ref{fig:Setup}(f). 

When rotating the phase mask to generate azimuthal polarization, the electron peak in the center fades away, as shown in Fig.~\ref{fig:Setup}(d). In this case, a significant amount of electrons is located on the other side of the hole, between the specular and the grazing directions. Overall, we remark that electrons are emitted in the specular direction only when radially polarized pulses, which possess a considerable longitudinal electric field, are used. This strongly suggests that these electrons are, as initially desired, accelerated by the $E_z$ field.

Figure~\ref{fig:spectra}(a) shows angularly resolved electron energy spectra for linear, radial and azimuthal polarization. We observe, in accordance with the angular distributions, that relativistic electrons are obtained in the specular direction only in the case of radial polarization. Figure~\ref{fig:spectra}(b) shows the electron energy spectra recorded at the position of the bright spot between the specular and normal directions for linear and azimuthal polarization and at the position of the specular spot for radial polarization. With linear polarization, the maximum reached energy is $11$\,MeV with a peak around $3$\,MeV while with radial polarization the maximum energy is $6$\,MeV with a peak around $2$\,MeV. For azimuthal polarization, the spectrum peaks at the same energy as for radial polarization but the maximum reached energy is smaller (around $4$\,MeV).

\begin{figure*}[ht!]
  \includegraphics[width=2.\columnwidth,scale=1]{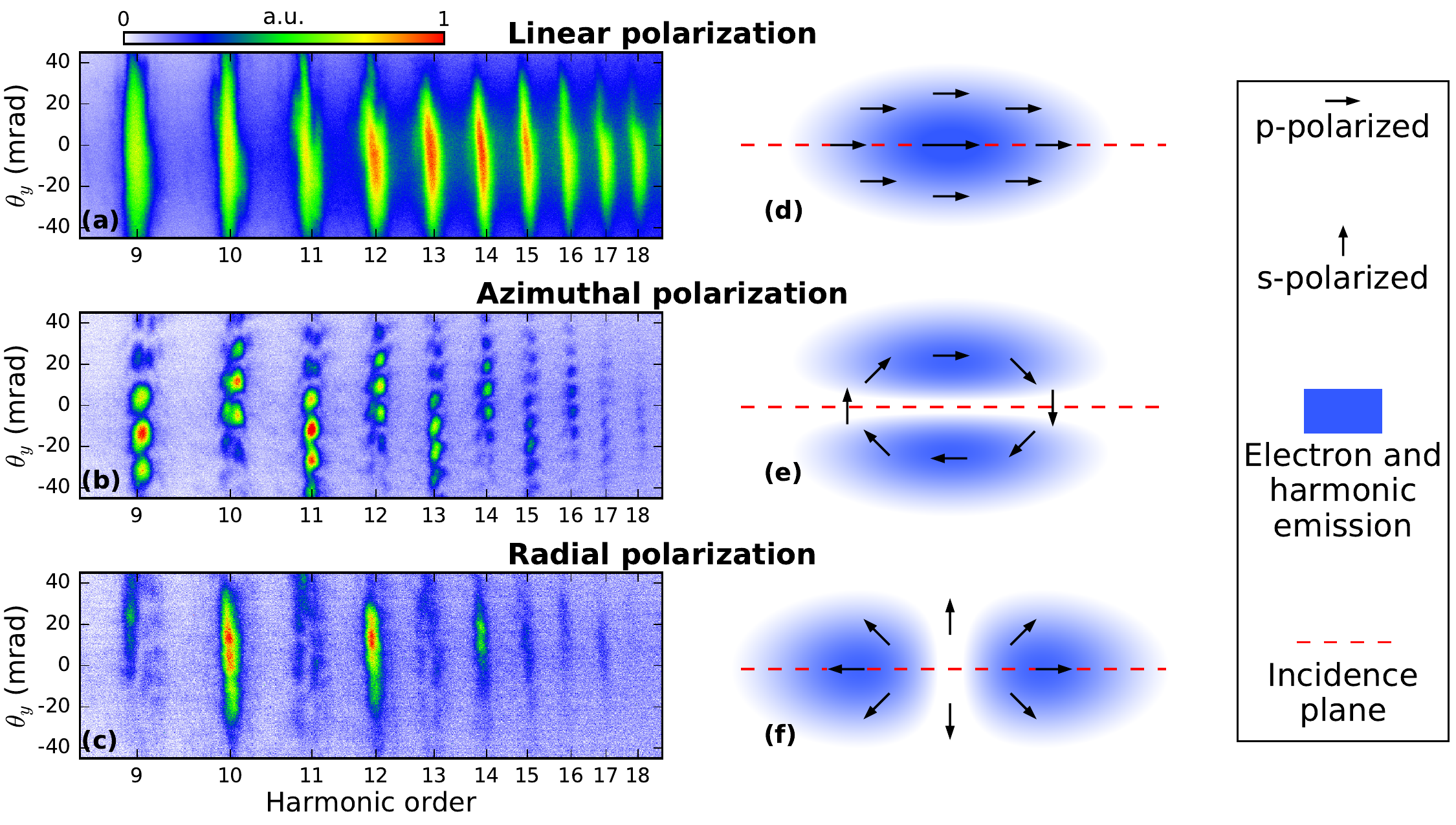}
  \caption{\label{fig:Principle} (a)-(c) Experimental angularly resolved harmonic spectra in the case of (a) linear polarization, (b) azimuthal polarization and (c) radial polarization. (d)-(f) Schematic view of the laser field footprint on the plasma surface for (d) linear, (e) azimuthal or (f) radial polarization. Electrons and harmonics are only emitted when the laser is close to p-polarized (i.e. when the electric field is almost parallel to the incidence plane), which results in two spatially-separated sources in the case of radial and azimuthal polarization.}
\end{figure*}

With the purpose of gaining additional insight on the interaction, we also study high-harmonic generation (HHG), which is closely related to the generation of fast electrons in the relativistic regime~\cite{Thevenet16_2,chop19}. Figures~\ref{fig:Principle}(a) to~\ref{fig:Principle}(c) show typical angularly resolved harmonic spectra for linear, azimuthal and radial polarization between harmonic orders 9 and 18. When the phase mask is introduced in the beam, the total harmonic yield decreases. Each spectrum is therefore renormalized separately in Fig.~\ref{fig:Principle}. In the case of azimuthal polarization, we observe interference fringes on each harmonics in the $\theta_y$ angle. These fringes correspond to the interference pattern that would be generated by two sources separated by $4.8 \pm 0.1\,\upmu$m in the transverse y-direction. The interference patterns for consecutive harmonic orders appear to be shifted by $\pi$. In other words, if the signal is close to 0 at a given angle for harmonic order $n$, it will be close to maximum at the same angle for harmonic order $n+1$. In the case of radial polarization, we observe no interference pattern but we notice that the harmonic intensity strongly varies between even and odd harmonics (in Fig.~\ref{fig:Principle}(c) the even harmonics are stronger but in some other shots the odd harmonics can become stronger).

\section{\label{sec:level4} Interpretation and discussion}

We analyze in this section the above-presented experimental results. We start by explaining in Sec.~\ref{subsec1} the observed patterns in the high-harmonic spectra and we discuss the resulting consequences for electron injection and for the structure of the reflected field. Based on these important findings, we then study in Sec.~\ref{subsection2} the acceleration of electrons in vacuum and the conditions that lead to narrow divergence electron beams in the specular direction.

\subsection{\label{subsec1} Harmonic generation and electron injection}
\subsubsection{Physical interpretation for the two sources of HHG}
The fact that the harmonic signal with an azimuthal polarization of the laser seems to originate from two sources from either side of the incidence plane strongly suggests that only the parts of the laser that are p-polarized (i.e. with an electric field component perpendicular to the plasma surface, which can drive the Relativistic Oscillating Mirror (ROM) mechanism) contribute to high-harmonic emission, while no harmonics are emitted in the regions where the laser is s-polarized. It has indeed been shown~\cite{chop19} that, at the laser intensities considered here, the harmonic signal is suppressed when the polarization is switched from p to s. These ideas are illustrated in Figs.~\ref{fig:Principle} (d) to (f): for linear polarization, the whole beam is p-polarized, which leads to a single harmonic source. On the other hand, for azimuthal or radial polarization, the locally p-polarized parts of the beam form two separate spots which can result in two separate harmonic sources. This is supported by the fact that the distance between the two sources deduced from the period of the interference pattern ($4.8 \pm 0.1\,\upmu$m) matches the distance between the two maxima in the focal spot ($4.3 \pm 0.2\,\upmu$m, see Fig.~\ref{fig:Setup}(b)).

This physical interpretation can also be used to explain why interference patterns for consecutive harmonic orders appear to be shifted by $\pi$. We indeed know from simulations that the harmonic signal is emitted once every optical period at a precise phase of the incident laser field in the form of attosecond pulses. Since the electric field in the two separate sources have opposite sign for radial or azimuthal polarization, the harmonic emission is delayed by half a laser cycle from one source to the other, i.e. a $\pi$ phase delay. This leads, for harmonic $n$, to a phase shift of $n\pi$ between the two sources, which results in a phase shift of $\pi$ between the interference patterns of consecutive harmonics.

For radial polarization, we also expect to see an interference pattern but this time along the x-direction (i.e. as a function of the $\theta_x$ angle). However, such fringes cannot be resolved in the experiment as our spectrometer samples the harmonic beam at a given $\theta_x$. Therefore, the spectrometer only sees one position in the interference pattern for each harmonic. Since there is a $\pi$ phase difference between consecutive harmonic patterns, the harmonic intensity can strongly vary between odd and even harmonics, which is precisely what we observe in Fig.~\ref{fig:Principle}(c).

\subsubsection{3D PIC simulations of harmonic generation}
In order to confirm this physical explanation, we turn to 3D PIC simulations using the code WARP~\cite{vay12,warp} combined with the high-performance library PICSAR~\cite{vinc17,picsar}. We use the high-order Pseudo-Spectral Analytical Time Domain (PSATD) Maxwell solver which strongly reduces numerical dispersion~\cite{vay13,vinc16}. As a result, convergence of the simulations is attained with larger spatial and temporal steps, which makes full 3D PIC simulations at solid density feasible using current petascale machines~\cite{blac17,vinc18}. Each of the simulations presented here costs around 3 million computing hours on the MIRA supercomputer~\cite{mira}. The gradient scale length is chosen to be close to the optimal value for electron ejection and high-harmonic generation and the laser parameters are the same as in experiments. A moving window is used to follow the trajectory of electrons in the reflected pulse up to 80 $\micron$ away from the plasma (approximately 2 Rayleigh lengths). More details regarding the simulations are given in the Supplemental Material~\cite{suppmat}.

\begin{figure}[ht!]
\centering
\includegraphics[width=1.\columnwidth,scale=1]{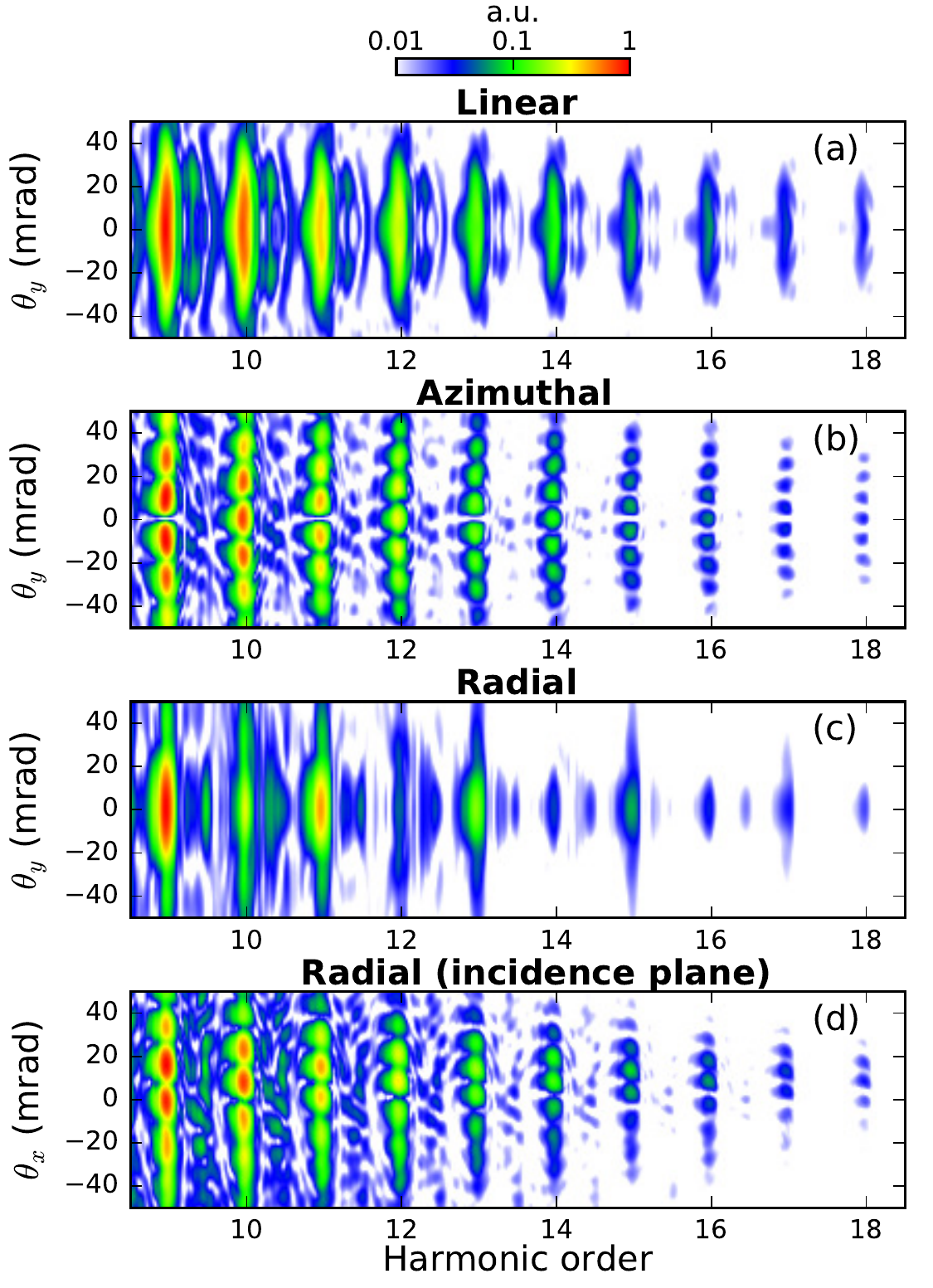}
\caption{\label{fig:HarmonicsPIC} Results from 3D PIC simulations. Angularly resolved harmonic spectra for (a) linear, (b) azimuthal or (c),(d) radial polarization. The angular dependence is shown with respect to the $\theta_y$ angle in panels (a) to (c), which corresponds to the experimental case, and with respect to the $\theta_x$ angle in panel (d), so that the interference pattern produced with a radially-polarized beam becomes apparent.}
\end{figure}

Angularly-resolved harmonic spectra obtained from the simulations are displayed in Fig.~\ref{fig:HarmonicsPIC}. In panels (a) to (c), the harmonics are resolved in the $\theta_y$ direction so that they can be directly compared with the experimental measurements shown in Fig.~\ref{fig:Principle}. A very good agreement is found: in azimuthal polarization, we observe interferences with a phase shift of $\pi$ between consecutive harmonics while in radial polarization, we notice that the harmonic intensity considerably differs between even and odd harmonics. In Fig.~\ref{fig:HarmonicsPIC}(d), the angular dependence is shown with respect to the $\theta_x$ angle in the radially polarized case, allowing us to uncover the interference pattern which is not visible in experiments. Both the experimental and simulated harmonic spectra show clear evidence that radial or azimuthal polarizations result in two separate harmonic sources that correspond to positions where the laser is locally p-polarized. This, again, is supported by the $\approx 4.7 \micron$ distance between the two sources obtained from the simulated harmonic signal, which is consistent with the theoretical distance between the two maxima in the focal spot ($\sqrt{2} w_0 \approx 4.4 \micron$).

\subsubsection{Consequences for electron injection}
As we have stated in the introduction, the harmonics observed in this interaction regime are emitted via the Relativistic Oscillating Mirror (ROM) mechanism, which occurs as the reflecting plasma surface oscillates nonlinearly following the laser field. At each laser optical cycle, when the oscillating electrons are pulled towards vacuum, the incident field is strongly Doppler upshifted, leading to the high-harmonic generation. At the same time, part of the oscillating electrons acquire a high enough energy to escape the plasma, leading to the ejection of electrons. Since both electron and high-harmonic emissions are originating from the same oscillation of the plasma surface, their emissions are synchronized and their yields are optimized for the same laser and plasma parameters. Such a correlation between ROM harmonics and electron ejection has indeed been observed experimentally~\cite{chop19}. In the case of radial or azimuthal polarization, we have just seen that harmonics are only generated at the positions where the laser is locally p-polarized. We can therefore expect a similar behavior for electron ejection, which would mean that electrons are also emitted from two separate sources.

\begin{figure}[t!]
\centering
\includegraphics[width=1.\columnwidth,scale=1]{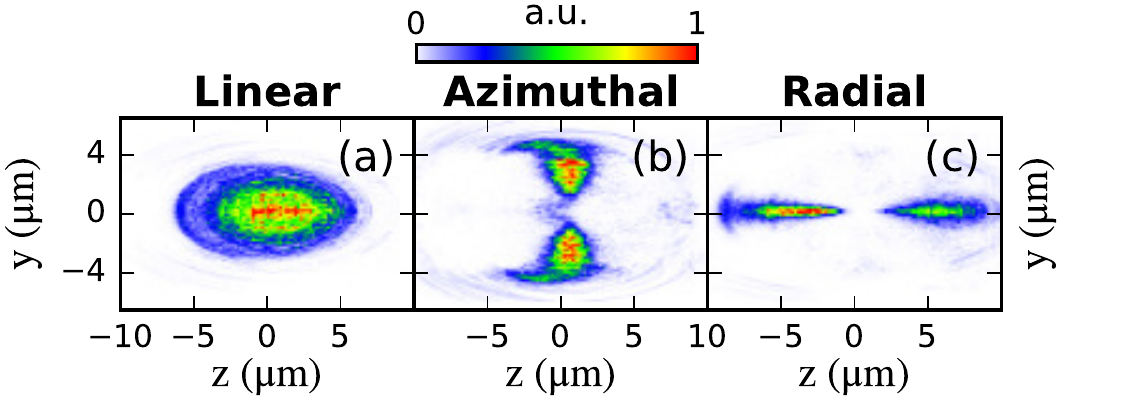}
\caption{\label{fig:InitPosPIC} Results from 3D PIC simulations. Initial position of the electrons that are ejected $4 \micron$ away from the plasma in the case of (a) linear, (b) azimuthal or (c) radial polarization.}
\end{figure}

In order to confirm this prediction, we plot in Fig.~\ref{fig:InitPosPIC} the initial position of the electrons that are ejected from the plasma in the 3D simulations. As anticipated, only the p-polarized parts of the laser contribute to electron injection, which results in two distinct electron sources in radial and azimuthal polarization. The fact that electron emission only occurs at specific parts of the focal spot can partly explain the significant shot-to-shot fluctuations observed in experiments with radial and azimuthal polarization. We have indeed seen that the focal spot possesses an asymmetry (see Fig.~\ref{fig:Setup}(b)) which can vary depending on the shots. Shots with a higher intensity in the p-polarized parts of the beam will result in a much higher detected charge than shots with a higher intensity in the s-polarized parts of the beam.

From a VLA point of view, the fact that the $E_z$ field in radial polarization does not contribute to electron injection and that the interaction is still dominated by the transverse fields is not ideal. Indeed, the initial purpose of using radial polarization was to accelerate the electrons close to the optical axis where the transverse fields are negligible whereas in our case the electrons are injected on the side of the beam where the transverse fields are the highest. Nevertheless, the fact that we have detected experimentally electrons in the longitudinal direction suggests that it is possible for some electrons to reach the optical axis after being injected on the side of the beam.

\subsubsection{Loss of radial symmetry of the reflected pulse}
\begin{figure}[ht]
\centering
\includegraphics[width=1.\columnwidth,scale=1]{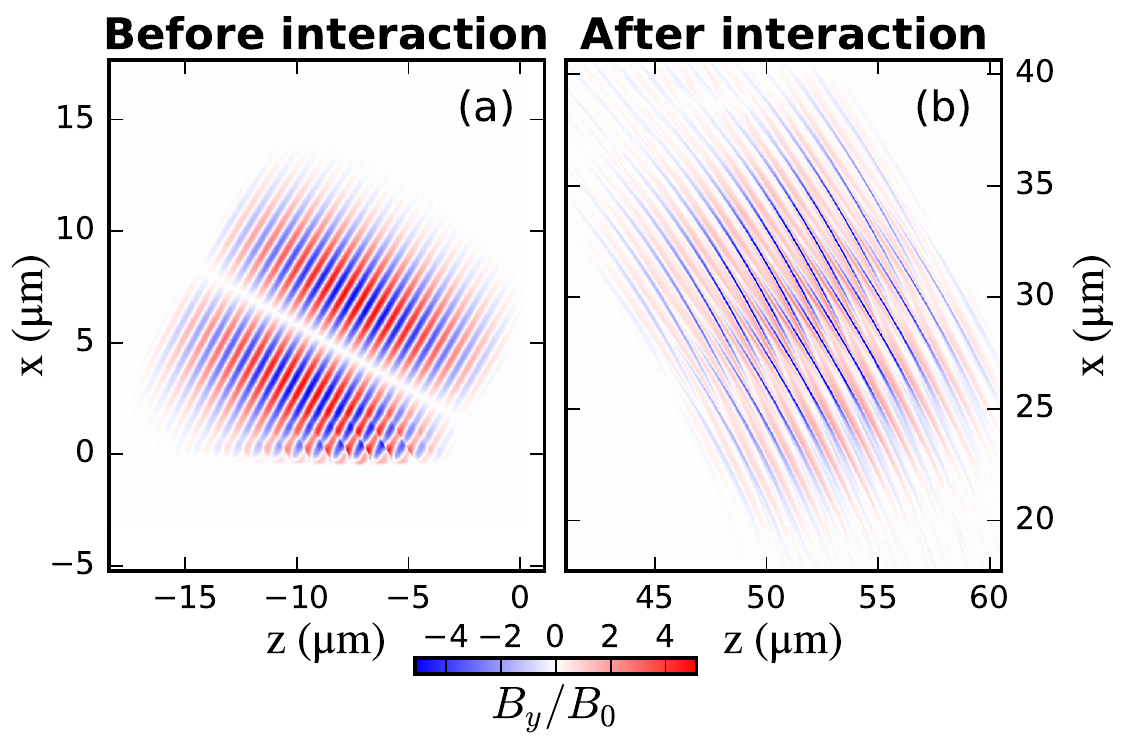}
\caption{\label{fig:Snapshot} Results from a 3D PIC simulation. Laser magnetic field in the incidence plane with radial polarization, either (a) $10\,\upmu$m before or (b) $60\,\upmu$m after reflection. Here, $B_0 = E_0/c$ with $E_0$ defined earlier.}
\end{figure}

Another important insight brought by the simulations is that the laser partially loses its spatial structure upon reflection. This is due to the fact that reflectivity of the laser fundamental frequency is lower in the p-polarized parts of the focal spot, and higher in the s-polarized parts of the spot~\cite{chop19} (this can be understood by considering the fact that electron emission and harmonic generation are stronger for the p-polarized components of the field). Consequently, the reflected laser field no longer has cylindrical symmetry. Figure~\ref{fig:Snapshot} displays snapshots from the PIC simulation with radial polarization showing the laser pulse in the incidence plane before and after reflection. We most notably observe that the hole in the intensity distribution at the center of the laser is suppressed after the interaction. This may make the VLA process in the reflected pulse more complex. In particular, we remark that high-amplitude harmonic fields, which have a much longer Rayleigh length than the fundamental pulse, remain present on the optical axis far from the plasma. Such on-axis harmonic fields could cause an unwanted deflection of electrons accelerated by the longitudinal field close to the optical axis, where transverse fields are supposed to be negligible in a perfect radially polarized pulse.

\subsection{\label{subsection2} Electron acceleration in vacuum}
\subsubsection{3D PIC simulations of electron acceleration}
The harmonic spectra obtained in PIC simulations with radial and azimuthal polarization show excellent agreement with experiments, but we find that it is more difficult to reproduce the electron angular and energy distributions. First, it must be noted that in order to obtain accurate results, the simulations should be run until electrons completely exit the laser field, which is a considerable computational challenge. In our simulations, the reflected pulse propagates $80 \micron$ away from the plasma, i.e. more than two Rayleigh lengths; nevertheless, some electrons are still interacting with the laser, indicating that their angular and energy distributions might still evolve. Secondly, as we have seen in the experiment, results obtained with the phase mask for radial polarization give lower performance in terms of electron acceleration (energy and charge) and harmonic yield, compared to the linear polarization case. This suggests that imperfections of the focal spot (such as intensity inhomogeneities and phase aberrations), which are not included in the PIC simulations, might affect the interaction more severely in the case of radial polarization, probably because the radial symmetry can easily be broken when imperfections are present in the focal region. 

\begin{figure}[t!]
\centering
\includegraphics[width=1.\columnwidth,scale=1]{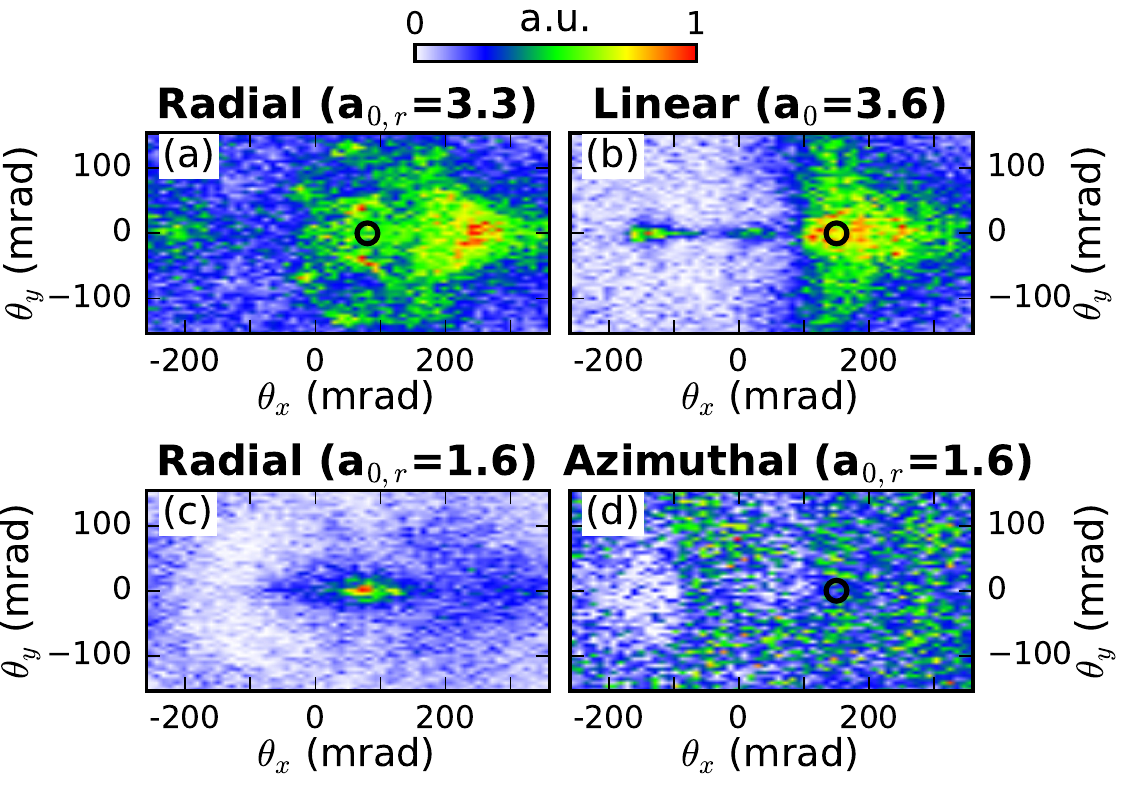}
\caption{\label{fig:AngularPIC} Results from 3D PIC simulations. (a)-(b) Angular distributions of the electrons with an energy greater than 1 MeV obtained at the end of the simulations with (a) radial polarization and $a_{0,r} =3.3$, (b) linear polarization and $a_{0} = 3.6$, (c) radial polarization and $a_{0,r} = 1.6$ and (d) azimuthal polarization and $a_{0,r} = 1.6$.}
\end{figure}

Therefore, we find that in order to obtain a quantitative agreement with the experiment, it is necessary to reduce the laser intensity substantially, and even more so in the case of radial polarization. Figures~\ref{fig:AngularPIC}(a) and \ref{fig:AngularPIC}(b) show typical electron angular distributions obtained in simulations with lower intensity for, respectively, radial and linear polarization. Here, $a_0=3.6$ for the linear case and $a_{0r}=3.3$ for the radial case, corresponding to a factor of two lower intensity compared to the experiment described in Sec.~\ref{sec:level2}.  In the linearly polarized case, we find that the main features of the experimental angular distribution are well reproduced: most electrons are accelerated between the specular and normal directions ($\theta_x > 0$) and there is virtually no electron emitted in the specular direction. This is in good agreement with previous 3D PIC simulations performed with similar laser and plasma parameters~\cite{vinc18,chop19}. When the polarization is switched to radial, there are more electrons located around the specular direction, as in experiments. However, we find that contrary to the experiment, this does not lead to a reduction of the electron beam angular spread and the overall simulated beam does not completely match the experimental results.

\begin{figure}[t!]
\centering
\includegraphics[width=1.\columnwidth,scale=1]{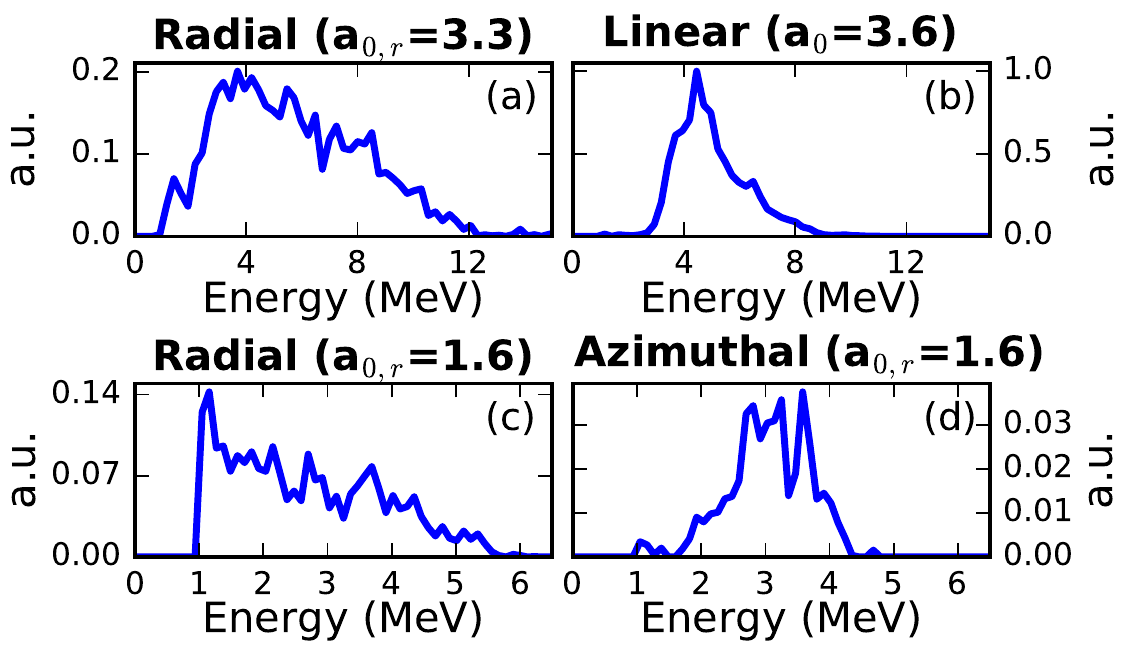}
\caption{\label{fig:SpectraPIC} Results from 3D PIC simulations. Electron energy distributions computed in a 30 mrad cone at the angle defined by the black circles shown in Fig.~\ref{fig:AngularPIC}. All four spectra are normalized in the same manner.}
\end{figure}

The corresponding energy spectra are computed at the angles indicated by the black circles and are shown in Figs.~\ref{fig:SpectraPIC}(a) and \ref{fig:SpectraPIC}(b). Even with this lower intensity level, the simulation with radial polarization yields electrons with an energy that is approximately twice as high as detected experimentally. This, again, suggests that imperfections of the laser focal spot degrade substantially the acceleration process compared to the ideal case that is considered in the simulations. This is consistent with the fact that both electron ejection from the plasma and subsequent VLA are sensitive to the exact spatio-temporal shape of the laser pulse, as was shown by recent theoretical studies on VLA~\cite{mart14,joll19}. 

To obtain an electron energy spectrum matching the experimental one, we have performed another simulation with radial polarization with an intensity even further decreased to approximately a tenth of the experimental value given in Sec.~\ref{sec:level2}. Although this is a large difference, we find that it leads to a good agreement with experiments, as can be seen in Fig.~\ref{fig:SpectraPIC}(c). This shows the extent to which the acceleration process seems to be negatively affected by the imperfections introduced by the phase mask. Interestingly, we find that reducing the laser intensity leads to a much more collimated electron beam, whose angular distribution is shown in Fig.~\ref{fig:AngularPIC}(c). This distribution resembles that of the best experimental shots with radial polarization (see Fig.~\ref{fig:Setup}(f)), which confirms that a better agreement with experiment is found by decreasing the intensity. We also once again find that using radial polarization can result in a significant decrease in the electron beam divergence.

Finally, simulations confirm that azimuthal polarization does not yield a collimated electron beam in the specular direction, in agreement with experiments. For example, in Fig.~\ref{fig:AngularPIC}(d), we show the angular distribution obtained with the same parameters as in Fig.~\ref{fig:AngularPIC}(c), except that the polarization is changed from radial to azimuthal. We find in this case that the electrons are emitted with a very wide divergence of the order of 600 mrad (which is not entirely visible in Fig.~\ref{fig:AngularPIC}(d), for direct comparison with the other numerical and experimental distributions). The corresponding energy spectrum displayed in Fig.~\ref{fig:SpectraPIC}(d) shows that, for the considered laser parameters, the energies reached with azimuthal polarization are slightly lower than those reached with radial polarization, in good agreement with experiments.

\subsubsection{Physical explanation for the narrow divergence e-beam}
\begin{figure}[ht!]
\centering
\includegraphics[width=1.\columnwidth,scale=1]{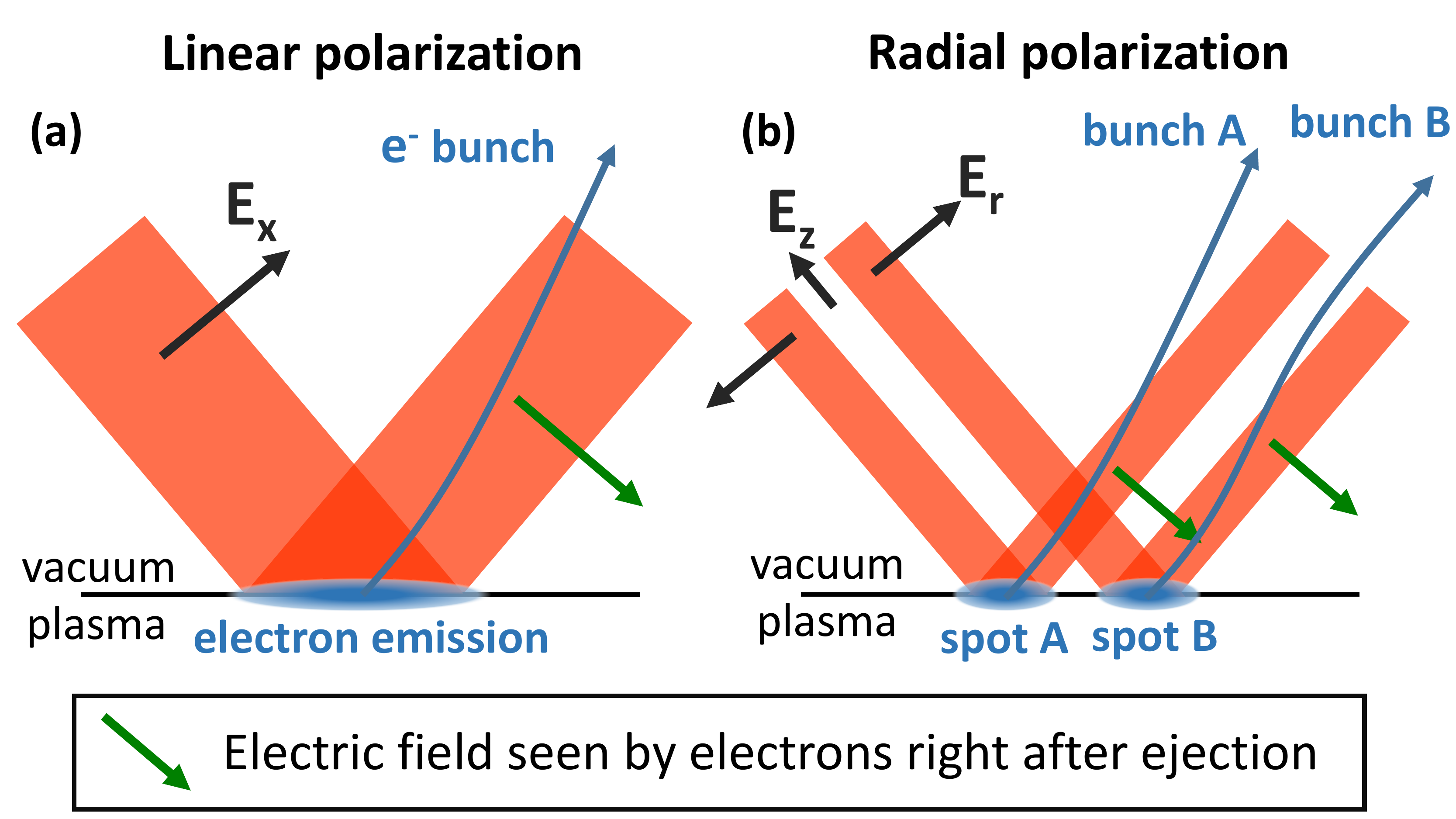}
\caption{\label{fig:EjectLinRad} Schematic illustration of electron emission from the plasma in the incident plane for linear (a) and radial (b) polarization. The black arrows represent the incident electric field. The green arrows in the reflected field indicate the direction of the transverse electric field seen by the electrons immediately after they are injected. The electrons are always emitted at a phase of the laser field such that they tend to be deviated towards the normal direction (this implies in particular that the time at which electrons are ejected is shifted by half a laser period between spot A and spot B). For radial polarization, the electrons originating from spot B are more likely to interact with the $E_z$ field than those coming from spot A.}
\end{figure}

In this section, we investigate the reasons why we obtain a narrow divergence electron beam with radial polarization when $a_{0,r} = 1.6$, but not when $a_{0,r} = 3.3$, with the ultimate purpose of understanding the conditions leading to a high-quality electron beam. In order to understand the fundamental differences between the two cases, we have analyzed the trajectories of the corresponding electrons in the PIC simulations. Recalling that electrons are injected in two spatially-separated bunches (corresponding to regions where the laser is locally p-polarized, as previously seen), we find that only the electrons originating from the bunch located on the right (such that $z > 0$) in Fig.~\ref{fig:InitPosPIC}(c) contribute to the collimated spot in the specular direction in Fig.~\ref{fig:AngularPIC}(c). This can be understood fairly easily and is illustrated in Fig.~\ref{fig:EjectLinRad}. First, we stress a general feature of this plasma mirror injection mechanism: most electrons are ejected from the plasma at a phase such that they tend to be deflected by the transverse fields towards the direction normal to the target surface, see Fig.~\ref{fig:EjectLinRad}(a). This explains why, in the linear polarization case, the electron beam is located between the specular and normal directions ($\theta_x > 0$). In the case of radial polarization, electrons initially in the bunch such that $z < 0$, labeled "bunch A" in Fig.~\ref{fig:EjectLinRad}, tend to directly escape on the side of the beam and thus never interact with the $E_z$ field. On the other hand, electrons initially in the bunch such that $z > 0$, labeled "bunch B" in Fig.~\ref{fig:EjectLinRad}, are shifted towards the center of the pulse, where they can be accelerated by the $E_z$ field. The fact that electrons have different trajectories depending on their initial location will also further increase the fluctuations in the experimental angular distributions when the focal spot possesses an asymmetry that varies from shot to shot, as was the case in the experiments.

\begin{figure}[ht!]
\centering
\includegraphics[width=1.\columnwidth,scale=1]{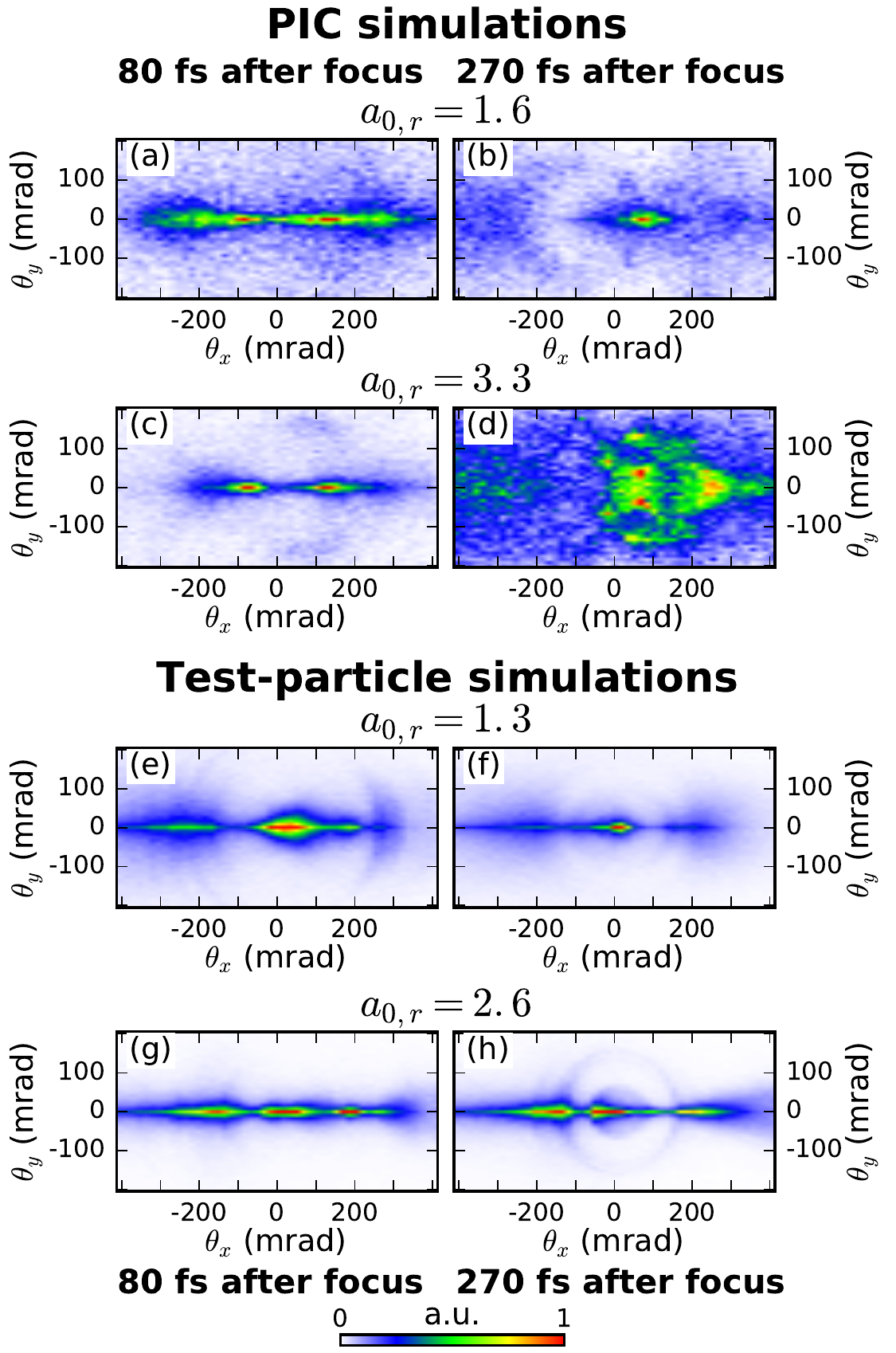}
\caption{\label{fig:ElecTraj} 3D-PIC simulations. (a)-(d) Angular distribution of electrons with $z > 0$ initially, with (a)-(b) $a_{0,r} = 1.6$ or (c)-(d) $a_{0,r} = 3.3$. The distributions are shown either $\approx 80$~fs after reflection or $\approx 270$~fs after reflection, which corresponds to the end of the simulation. (e)-(h) Similar angular distributions reproduced in test-particle simulations. The laser intensity is slightly reduced in the test-particle simulations to take into account the energy absorption by the plasma mirror.}
\end{figure}

We now focus on the electrons initially in "bunch B", both in the simulation with $a_{0,r} = 1.6$ and the simulation with $a_{0,r} = 3.3$. In conformity with the schematic view of Fig.~\ref{fig:EjectLinRad}(b), we observe in the PIC simulations that these electrons, initially on the side of the reflected laser pulse, propagate towards the center of the beam and subsequently spend a long time close to the optical axis where the longitudinal field has a high amplitude. We show in Figs.~\ref{fig:ElecTraj}(a) to (d) the angular distribution of the electrons initially in "bunch B", either $\approx 80$~fs after reflection or at the end of the simulation, $\approx 270$~fs after reflection. At the beginning of their interaction, these electrons have a very narrow angular spread in the $\theta_y$ direction in both simulations. In the lower intensity case ($a_{0,r} = 1.6$), this angular divergence remains small until the end, and we eventually observe a collimated beam in the specular direction. On the opposite, when the intensity is increased, a considerable widening of the angular distribution in the $\theta_y$ direction is visible. This means that electrons initially accelerated in the longitudinal direction are deflected in the transverse direction during their interaction with the reflected beam. This deflection, occurring far from the plasma, could be due to the harmonic fields which, as previously stated, have a longer Rayleigh length and remain for a long time close to the optical axis. The harmonic fields are much more intense in the simulation with $a_{0,r} = 3.3$ because the high-harmonic generation efficiency increases nonlinearly with the intensity~\cite{Thaury10}, which would explain why the electrons are not deflected in the simulation with $a_{0,r} = 1.6$.

\subsubsection{Test particle simulation of electron acceleration}
In order to confirm the deleterious role of high-harmonic fields, we have performed a series of fast 3D test particle simulations of the acceleration of an electron bunch by an ideal radially polarized pulse, without high-harmonic fields. To simplify the interaction, we only consider a single attosecond electron bunch which is, as in experiments, initially located off the optical axis where the transverse fields are the most intense. The electrons in this bunch start their interaction at a zero of the $E_r$ field, which corresponds to injection by a plasma mirror. At the beginning of the simulation, all electrons are on the same side of the focal spot, which corresponds to "spot B" in Fig.~\ref{fig:EjectLinRad}(b). This means that they are initially pushed by the transverse fields towards the optical axis. Electrons are injected at the focus of the laser, around four optical cycles before the temporal center of the pulse and have a Gaussian distribution both in real and momentum space, with an adjustable mean and variance. More details concerning these simulations are provided in the Supplemental Material~\cite{suppmat}. 

By choosing the adequate initial electron distribution, we find that it is possible to obtain similar trajectories as in the PIC simulation with the lowest intensity. For example, we show in Figs.~\ref{fig:ElecTraj}(e) and (f) the angular distributions obtained either $\approx 80$~fs after focus or $\approx 270$~fs after focus for an electron beam initially located $2 \micron$ from the optical axis with an initial mean kinetic energy of 1 MeV and an average angle of $10^{\circ}$ away from the specular direction, oriented towards the grazing direction. The initial standard deviations of the Gaussian distributions are $\sigma_{z} = 0.17 \micron$ and $\sigma_{pz} = 0.9~ m_e c$ in the longitudinal direction and $\sigma_{x} = 0.5 \micron$, $\sigma_{y} = 1.2 \micron$, $\sigma_{px} = m_e c$ and $\sigma_{py} = 0$ in the transverse directions. We observe as in the PIC simulations the formation of a collimated spot in the specular direction.

The fact that the test-particle simulation can reproduce the lowest intensity 3D PIC simulation - and thus the experimental results - is particularly valuable because it allows us to extract useful information regarding the trajectories of the accelerated electrons. For example, we can calculate the work done by the different components of the electric field and also evaluate the amount of energy gained \textit{in vacuum} by the electrons (as opposed to the energy gained from plasma fields during the laser-plasma interaction). As expected, we find that the electrons emitted in the collimated spot gain their energy from the work done by the $E_z$ field. We can additionally estimate that these electrons exit the plasma with an energy of the order of 1~MeV and are accelerated in vacuum by the radially polarized beam to energies ranging between 2~MeV and 5~MeV. This corresponds to a maximum energy gain in vacuum of approximately 4~MeV. 

When increasing the intensity in the test-particule simulations, we do not obtain the large broadening of the angular spread in the $\theta_y$ direction that was observed in the PIC simulation. For instance, we show in Figs.~\ref{fig:ElecTraj}(g) and (h) the angular distributions obtained for an electron beam initially located $3.1 \micron$ from the optical axis with an initial mean kinetic energy of 2 MeV oriented towards the specular direction. The initial standard deviations of the Gaussian distributions are $\sigma_{z} = 0.19 \micron$ and $\sigma_{pz} = m_e c$ in the longitudinal direction and $\sigma_{x} = 0.8 \micron$, $\sigma_{y} = 0.8 \micron$, $\sigma_{px} = 3.2~m_e c$ and $\sigma_{py} = 0$ in the transverse directions. We observe that the electron beam remains collimated in the y-direction, which strongly differs from the PIC simulation. This is an indication that the high-harmonics fields, which are not included in the test-particle simulations, are indeed deflecting the electrons off the optical axis and are detrimental to the electron beam quality. \\

Overall, the 3D PIC and test-particle simulations confirm that radial polarization can lead to acceleration in the longitudinal direction, driven by the $E_z$ field. This can lead to a decrease in the divergence of the accelerated electrons compared to linear polarization. However, the simulations also indicate that the experiments were not performed with ideal interaction conditions. First, the electrons are mainly injected by the $E_r$ field at a position of the beam where the longitudinal field is negligible. Secondly, the radially polarized structure is partially lost upon reflection. This results in particular in strong harmonic fields which may deviate the electrons located on the optical axis if the laser is intense enough. These limitations make this acceleration method difficult to scale to larger laser and electron energies because we expect in this case an increase in the electron beam angular spread induced by the high-harmonics. 

\section{\label{sec:level5} Ideal physical conditions for VLA with radial polarization}

In this section, we explain how to overcome the previously mentioned limitations. To this end, we describe the interaction parameters that should be used to take full advantage of radial polarization while using similar laser pulse energies. In particular, two key elements can be changed in order to drastically improve the results.

The first one is the use of normal incidence, as was previously done in Ref.~\cite{Zaim17} for few-mJ pulses, rather than oblique incidence. With normal incidence, the $E_r$ field is parallel to the plasma surface so that only the $E_z$ field contribute to electron injection. Electrons are therefore injected close to the optical axis where the longitudinal field can accelerate them efficiently. This is a considerable improvement compared with oblique incidence where the electrons are injected in regions with intense radial fields. Moreover, at normal incidence high-harmonic generation is strongly reduced and the axi-symmetry of the reflected pulse is preserved so that the transverse fields will always vanish on axis. This means that electrons will not be deflected from the region with strong accelerating fields.

It should be mentioned that using normal incidence would lead to additional experimental challenges since in this case the laser pulse is reflected back into the laser system and the accelerated electrons propagate towards the focusing parabola. Fortunately, there are solutions to overcome these issues: a Faraday rotator can be used to prevent the reflected laser pulse from causing damage to the laser system and the accelerated electron beam can be collected by inserting a hole in the center of the parabola, where the incident laser intensity is negligible in the case of radial polarization. Alternately, these challenges may be avoided altogether by using small incidence angles rather than complete normal incidence.

The second key element is the increase of the amplitude of the longitudinal field $a_{0,z}$. This can be done both by focusing the laser pulse more tightly, as $a_{0,z}$ scales as $1/w_0^2$, and by reducing the pulse duration in order to increase the laser power. Using shorter pulses has other advantages: it leads to shorter electron beams and limits the interaction in the interference pattern formed by the incident and reflected beam. 

\begin{figure}[ht]
\centering
\includegraphics[width=1.\columnwidth,scale=1]{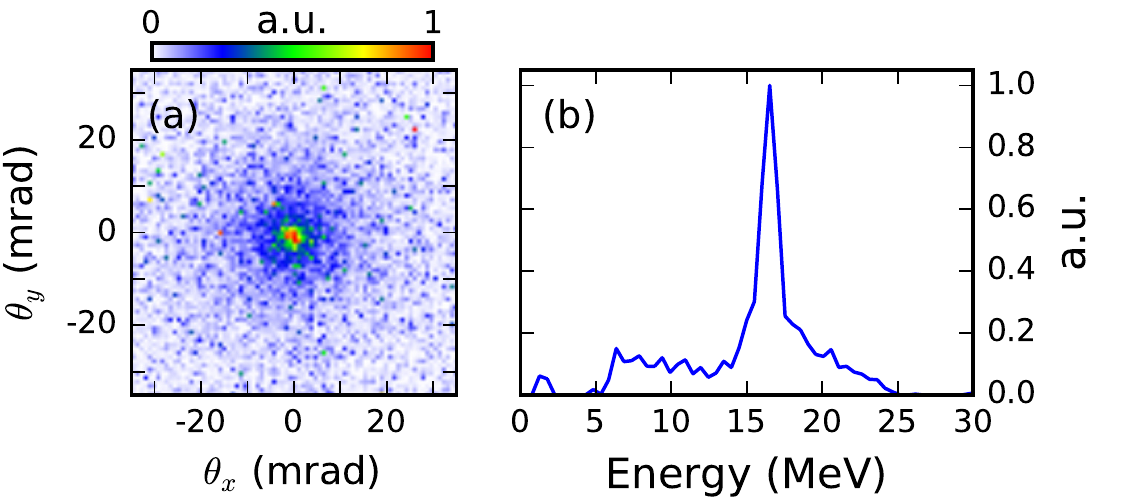}
\caption{\label{fig:Calder} Results from the PIC simulation at normal incidence. (a) Angular distribution of the electrons with an energy greater than 1 MeV around the specular direction. (b) Energy spectrum of the electrons accelerated within 5 mrad of the specular direction.}
\end{figure}

Figure~\ref{fig:Calder} presents results from a PIC simulation carried out in these conditions. Since the interaction at normal incidence has axial symmetry, the simulation is performed in cylindrical coordinates with the code CALDER-CIRC~\cite{lifs09}, thus hugely reducing its numerical cost. We use parameters targeted by the SYLOS laser system~\cite{kuhn17}: a pulse energy of 100 mJ, a beam waist of $w_0 = 1.5 \micron$ and a pulse duration of 5 fs in FWHM of intensity. This leads to $a_{0,z} = 3.78$ and $a_{0,r} = 9.55$. See also Supplemental Material~\cite{suppmat} for more information regarding this simulation. We observe that the charge emitted within an angle $\theta$ of the specular direction scales quasi-linearly with $\theta$. For instance, there is an emitted charge of 23.7 pC within 100 mrad of the specular direction, 9.84 pC within 50 mrad and 1.45 pC within 10 mrad. This means that the emitted charge per solid angle scales as $1/\theta$, which results in the peaked angular distribution shown in Fig.~\ref{fig:Calder}(a). The energy spectrum of the electrons accelerated in the specular direction, visible in Fig.~\ref{fig:Calder}(b), presents a narrow peak around 16 MeV. Such a narrow divergence ultrashort relativistic beam with quasi-monoenergetic energy could be particularly useful for applications such as ultrafast electron diffraction and femtosecond X-ray generation via inverse Compton scattering.

\section{\label{sec:level6} Conclusion}
We have demonstrated for the first time the possibility to accelerate electrons to relativistic energies with a radially polarized laser pulse. It was shown that radial polarization can lead to acceleration in the longitudinal direction directly from the $E_z$ component of the electric field and reduce the divergence of the electron beam. This work clearly demonstrates that the relativistic regime is reached for radially polarized laser pulses, with energy gains in vacuum reaching up to 4~MeV. However, the data shows that the performance is still degraded when compared to linear polarization. This suggests that more work is still required to generate higher quality radially polarized pulses at ultrahigh intensity. We have also unveiled the details of the physics of the laser-plasma interaction, and found that the use of radial or azimuthal polarization at oblique incidence leads to two spatially separated sources of electrons and high-harmonics. We conclude that harmonic generation is detrimental to the acceleration process so that these experiments were not performed in ideal conditions because of the large incidence angles used. We expect that the accelerated electron beams can be greatly improved at normal or quasi-normal incidence and with stronger longitudinal fields. These results may provide a new path for generating high quality ultra-short relativistic electron bunches. 

\begin{acknowledgments}
The authors thank Dr. G\"{o}tz Zinner from Bernard Halle (Germany) for producing the phase mask and Guillaume Blaclard for fruitful discussions. The research leading to these results has received funding from LASERLAB-EUROPE (grant agreement No. 654148, European Union’s Horizon 2020 research and innovation programme), the Agence Nationale pour la Recherche (Contract No. ANR-14-CE32-0011-03 APERO), the European Research Council (Contracts No. 306708, ERC Starting Grant FEMTOELEC and No. 694596, Grant ExCoMet), the Swedish Research Council and the Knut and Alice Wallenberg Foundation. An award of computer time (PlasmInSilico) was provided by the Innovative and Novel Computational Impact on Theory and Experiment (INCITE) program. This research used resources of the Argonne Leadership Computing Facility (MIRA), which is a U.S. DOE Office of Science User Facility supported under Contract No. DE-AC02-06CH11357.
\end{acknowledgments}

\end{document}